\documentclass[twocolumn,twoside,slac_two]{revtex4}
\usepackage{graphicx}
\usepackage{fancyhdr}
\usepackage{multirow}
\newcommand{\MET}{\mbox{$E\kern-0.60em\raise0.10ex\hbox{/}_{T}$}}
\newcommand{\MetVec}{\mbox{$\vec E\kern-0.60em\raise0.10ex\hbox{/}_{T}$}}
\newcommand{\PtVec}{\mbox{$\vec p\kern-0.45em\hbox{/}_{T}$}}
\newcommand{\lnqq}{\ell \bar{\nu} q \bar{q}}
\newcommand{\nnqq}{\nu \bar{\nu} q \bar{q}}
\newcommand{\ppbar}{p \bar{p}}
\newcommand{\bbbar}{b \bar{b}}

\newcommand{\MetDeltaPhi}{\Delta\phi_{\MET}^{\rm jet}}
\newcommand{\TrackDeltaPhi}{\Delta\phi(\MetVec,\PtVec)}

\begin{document}

\title{First Observation of Diboson Production in Hadronic Final State at the Tevatron}

%

\author{J.~Pursley, on behalf of the CDF Collaboration}
\affiliation{Department of Physics, University of Wisconsin-Madison, Madison, WI, 53706, USA}

\begin{abstract}
We present the first observation in hadronic collisions of the electroweak production of 
vector boson pairs ($VV$; $V=W,Z$) where one boson decays to a hadronic final state. The data 
correspond to 3.5 fb$^{-1}$ of integrated luminosity collected by the CDF II detector in 
$\ppbar$ collisions at a center-of-mass energy of 1.96 TeV. 
Event selection requires two identified jets and large transverse momentum imbalance. 
The analysis employs several novel techniques to 
suppress multijet background and reduce systematic uncertainties. We observe 
$1516\pm239$(stat) $\pm144$(syst) diboson candidate events and measure a cross section 
of $\sigma(\ppbar \to VV+X) = 18.0\pm2.8{\rm (stat)}\pm2.4{\rm (syst)}\pm1.1{\rm (lumi)}$
pb, in agreement with standard model (SM) expectations.
\end{abstract}

\maketitle

\thispagestyle{fancy}


\section{Introduction\label{sec:intro}}
The production of heavy gauge boson pairs ($WW$, $WZ$, and $ZZ$) has been observed in 
fully leptonic final states at the Fermilab Tevatron collider \cite{CDFlep,D0lep}.  
Evidence for diboson decays into an $\lnqq$ final state, with
$\ell = e, \mu, \tau$ and $q = u, d, s, c, b$, was recently presented by the D0
collaboration \cite{D0had}.  The analysis presented here is the first conclusive
observation at a hadron collider of diboson production with one boson decaying
into leptons and the other into hadrons \cite{CDFhad}.

Measurements of the diboson production cross sections provide tests of the
self-interactions of the gauge bosons, and deviations from SM predictions could indicate 
new physics \cite{Hagiwara:1986vm}.  Diboson production involving hadronic decays
is also topologically similar to associated Higgs boson production, $\ppbar \to VH + X$,
when the Higgs boson decays to $\bbbar$, the most promising discovery channel for a 
low-mass Higgs boson.  Thus the analysis techniques described here will also be useful 
for Higgs boson searches.

\section{Experimental Apparatus\label{sec:detector}}
The CDF II detector is described in detail in Ref.~\cite{CDFII}.  Surrounding the
beam pipe is a tracking system consisting of a silicon microstrip detector, a
cylindrical drift chamber, and a solenoid producing a 1.4T magnetic field along
the beam axis.  The central and forward calorimeters surround the tracking
system with a projective tower geometry.  The calorimeters are composed of inner
electromagnetic and outer hadronic sections that consist of lead-scintillator
and iron-scintillator respectively.  A system of Cerenkov counters, located 
around the beam pipe and inside the forward calorimeters, measures the number of
inelastic $\ppbar$ collisions per bunch crossing and thus the luminosity \cite{Acosta:2002hx}.

The geometry of the detector is characterized by the azimuthal angle $\phi$ and the 
pseudorapidity $\eta = -\ln [ \tan (\theta/2) ]$, where $\theta$ is the polar 
angle measured from the proton beam direction.
The transverse energy $E_T=E\sin\theta$, where $E$ is the energy in the calorimeter 
towers associated with a cluster of energy deposition.  Transverse momentum $p_T$ 
is the track momentum component transverse to the beam-line.
The missing transverse energy vector $\MetVec$ is defined as $-\sum_{i} E_T^i ~ \hat{n}_T^i$, 
where $\hat{n}_T^i$ is the transverse component of the unit vector pointing from the 
interaction point to the energy deposition in calorimeter tower $i$.   
This is corrected for the $p_T$ of muons, which do not deposit all of their energy in the 
calorimeter, and tracks which point to uninstrumented regions in the calorimeter.  
The missing transverse energy $\MET$ is defined as $|\MetVec|$.
Strongly interacting partons undergo fragmentation that results in highly collimated 
jets of hadronic particles.  Jet candidates are reconstructed in the calorimeter using
the {\sc jetclu} cone algorithm \cite{jetclu} with a cone radius of 0.4 in ($\eta,\phi$) 
space.  The energy measured by the calorimeter must be corrected to improve the 
estimated energy~\cite{JES}.  The effects corrected for include the non-linear response
of the calorimeter to particle energy, the loss of energy in uninstrumented regions of the
detector, and the energy radiated outside of the jet cone.

\section{Event Selection\label{sec:selection}}
Events are selected which have large $\MET$ and exactly two jets whose 
invariant mass can be reconstructed.  This signature is sensitive to both $\lnqq$ and 
$\nnqq$ decays because a charged lepton is not explicitly required in the final state.  
Due to the limited dijet mass resolution, there is significant overlap between the 
$W \to q \bar{q}'$ and $Z \to q \bar{q}$ dijet mass peaks.  Therefore we consider
as signal the combination of three diboson signals ($WW$, $WZ$, and $ZZ$), and search 
for diboson production in the dijet mass range $40 < M_{jj} < 160$ GeV$/c^2$.

Events are selected using a set of $\MET$-based triggers with a variety of $\MET$ and
jet requirements.  All of these triggers have benefited significantly from a calorimeter
trigger upgrade completed in 2007 \cite{trigger}.  The majority (94\%) of events satisfy
the inclusive $\MET$ trigger which requires $\MET > 45$ GeV.  We require events to have
$\MET > 60$ GeV and exactly two jets with $E_T > 25$ GeV and $|\eta| < 2.0$, which ensures 
a trigger efficiency of $96\% \pm 2\%$ on signal as measured in $Z \to \mu\mu$ events.  
Additionally, the fraction of the total
jet energy which is in the electromagnetic calorimeter is required to be less than 90\% to 
eliminate the possibility that electrons and photons are reconstructed as jets.
Several criteria suppress contamination from non-collision backgrounds.  
Events are required to have at least one reconstructed vertex formed by charged particle 
tracks to remove cosmic-ray events.  To reduce beam-related backgrounds, the electromagnetic
fraction of the total event energy must be greater than 30\%.  Additionally, the arrival
time of the jets as measured by the electromagnetic shower timing system \cite{timing}
must be consistent with the $\ppbar$ collision time.  After all selection criteria are 
made, non-collision backgrounds are a negligible contribution to the sample (fewer than 
90 events out of the total 44,910 selected events).

\begin{figure}
\includegraphics[width=80mm]{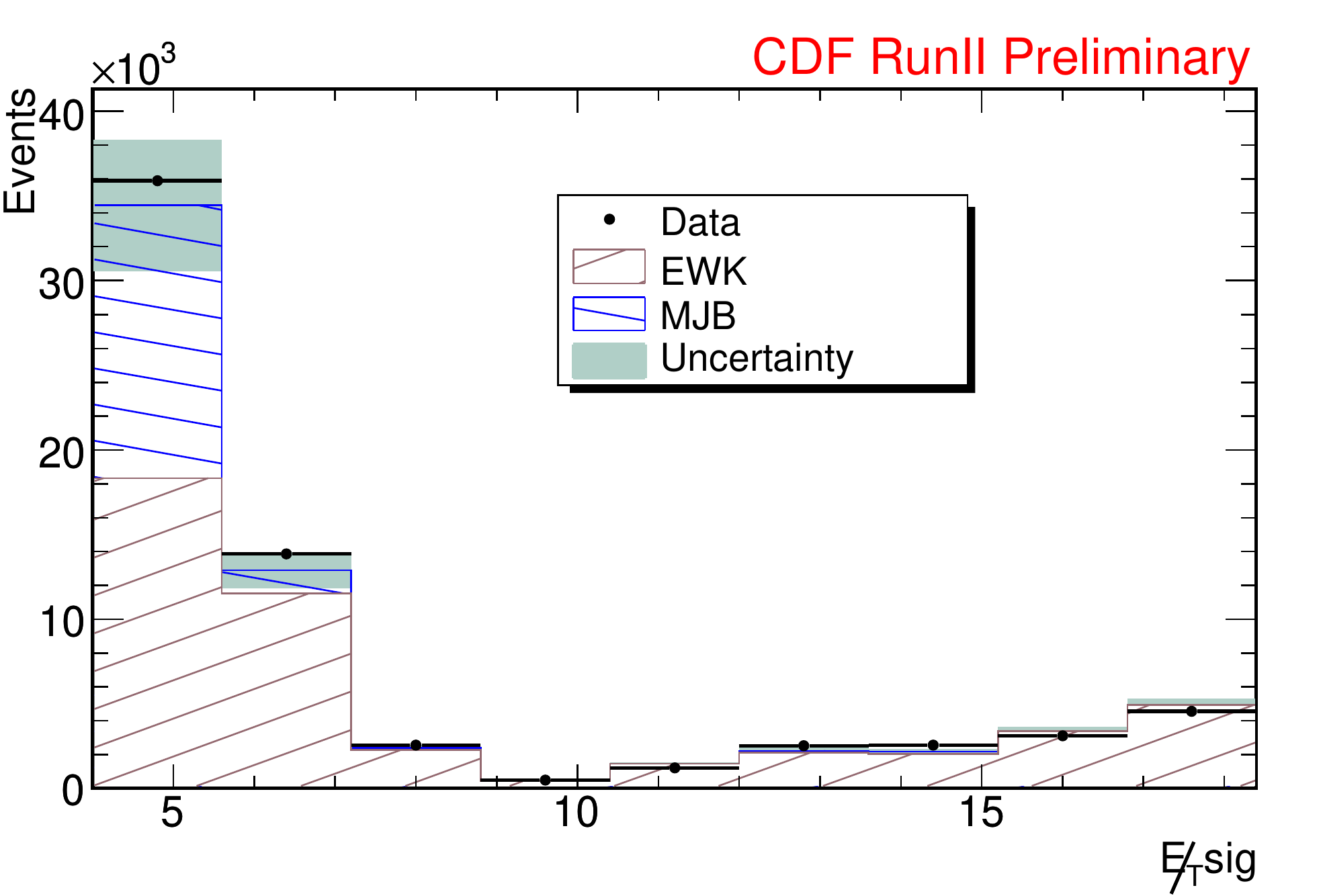}
\includegraphics[width=80mm]{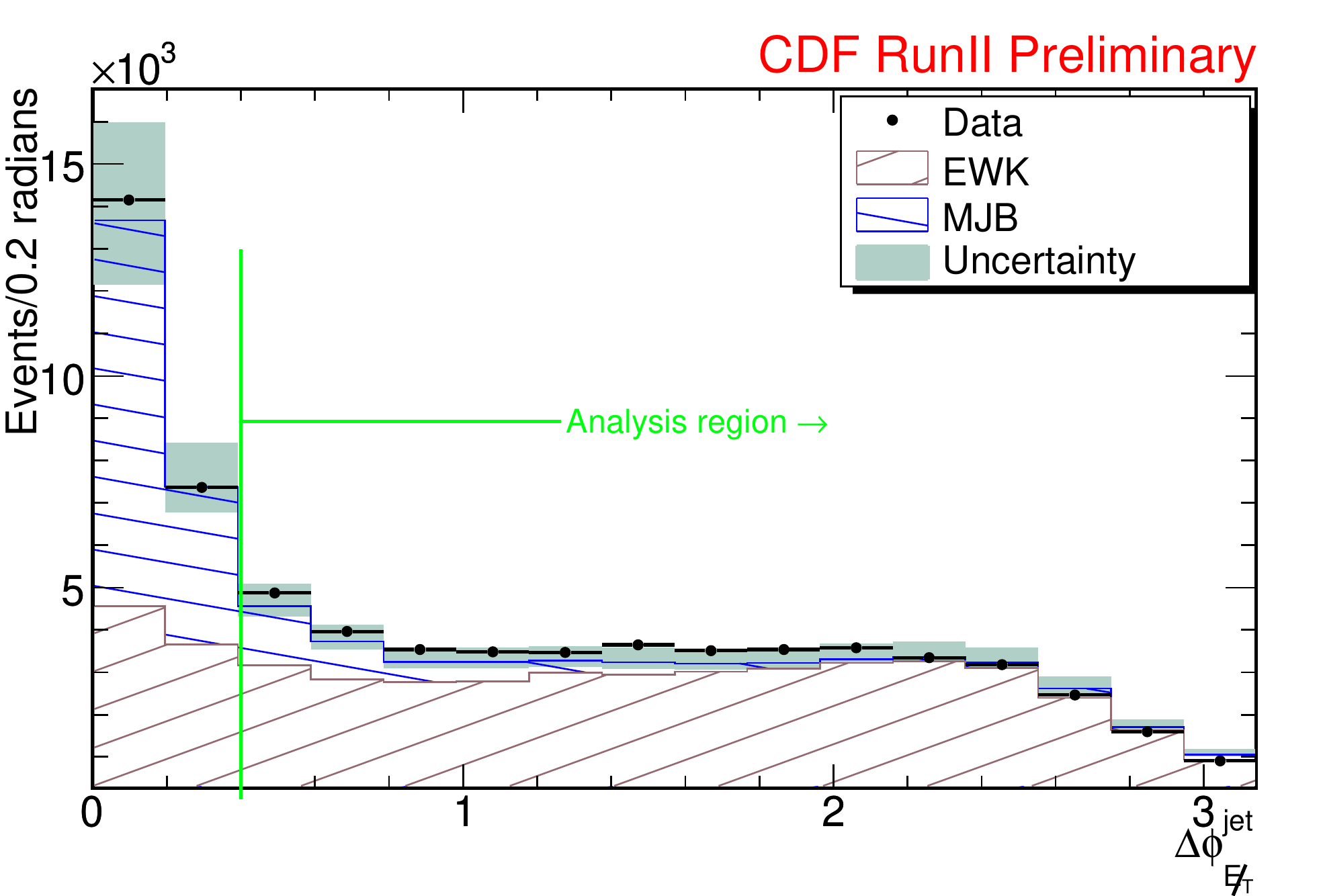}
\caption{Data compared with the sum of the predicted electroweak (EWK) and 
	multijet background (MJB) contributions for the $\MET$ significance 
	(top) and $\MetDeltaPhi$ (bottom) variables defined in Sec.~\ref{sec:selection}.  
	The expected signal is included here in the EWK contribution.  The 
	gray bands represent the total systematic uncertainty described in 
	Sec.~\ref{sec:syst}.  To reject events with fake $\MET$ we require 
	$\MET$ significance $> 4$ and $\MetDeltaPhi > 0.4$ radians.\label{fig1}}
\end{figure}

In order to suppress the multijet background described in Sec.~\ref{sec:mjb}, we use a
$\MET$ resolution model to distinguish true $\MET$ originating from neutrinos from fake 
$\MET$ arising from mismeasurement of jet energies.  The $\MET$ significance is a
dimensionless quantity based on the event topology, the energy resolution of the jets, 
and the soft unclustered particles.  The $\MET$ significance is calculated as follows:
\begin{eqnarray}
\MET~{\rm significance} = -\log_{10}(\bar{\mathcal{P}}), \nonumber  \\
{\rm where}~\bar{\mathcal{P}} = \prod \int_{-1}^{y_i} \mathcal{P}_i(x)dx~{\rm if}~y_i < 0  \nonumber \\
{\rm or}~\bar{\mathcal{P}} = \prod \left( 1 - \int_{-1}^{y_i} \mathcal{P}_i(x)dx \right)~{\rm if}~y_i > 0, \nonumber \\
{\rm with}~y_i = \MET/(E_T^i \cos\Delta\phi_i).
\label{eq-metsig}
\end{eqnarray}
In Eq.~\ref{eq-metsig}, $\mathcal{P}_i(x)$ is the jet energy resolution function of the 
$i$-th jet, $E_T^i$ is the energy of the $i$-th jet, and $\Delta\phi_i$ is the azimuthal 
angle between the $i$-th jet and $\MET$.
The jet energy resolution function $\mathcal{P}(x)$, defined as a ratio of the
detector-level and hadron-level jet energy, is obtained from dijet {\sc pythia}
Monte Carlo \cite{pythia} as a function of a jet's energy and pseudorapidity.  The 
function is validated in inclusive $Z \to ee$ events.
The $\MET$ significance is typically low when $\MET$ arises from mismeasurement.  In
addition to having a small significance, the $\MetVec$ will often be aligned with a jet.
Distributions of the $\MET$ significance and azimuthal angle between the $\MetVec$ and 
nearest jet ($\MetDeltaPhi$) after all previously described selection criteria are shown 
in Figure \ref{fig1}.  To reduce the multijet background, we select events with $\MET$ 
significance larger than 4 and $\MetDeltaPhi$ greater than 0.4 radians.

%


\section{Data Modeling\label{sec:model}}
The diboson signals ($WW$, $WZ$, and $ZZ$) are simulated by the {\sc pythia}
Monte Carlo generator.  After selection, the most significant
backgrounds to the diboson signal are multijet production and electroweak (EWK) 
processes such as $W(\to \ell \bar{\nu})$+jets and $Z(\to \nu \bar{\nu})$+jets.  
Less significant EWK backgrounds include $Z(\to \ell\ell)$+jets, top-quark pair
production, and single top-quark production.
The geometric and kinematic acceptance for all electroweak processes are determined 
using a Monte Carlo calculation of the collision followed by a 
{\sc geant3}-based simulation of the CDF II detector response \cite{GEANT}. 
Modeling of the backgrounds is described in more detail in the following sections.

\subsection{Electroweak Background\label{sec:ewk}}
The electroweak (EWK) processes considered to be backgrounds to this measurement
are the $V$+jet and top-quark production processes.
The $W(\to \ell \bar{\nu})$+jets processes are simulated by the
fixed-order matrix element generator {\sc alpgen} \cite{alpgen}
interfaced with {\sc pythia} to simulate parton showering and
fragmentation, the underlying event, and additional $\ppbar$
interactions in the event.
The $Z$+jets and top-quark production are simulated by {\sc pythia}.
The expected yields of all Monte Carlo-simulated background processes are 
normalized to SM cross sections calculated at next-to-leading order,
although the overall normalization of this background is allowed to float 
in the final fit to data.

\subsection{Multijet Background\label{sec:mjb}}
The multijet background (MJB) does not typically result in signatures of large 
$\MET$.  However, when jet energy is not measured accurately an event may be 
reconstructed with large fake $\MET$ and thus pass the event selection.  Due
to the large rate of MJB production, this is still a significant background.
Much of the MJB is removed by the requirements on the $\MET$ significance and
$\MetDeltaPhi$ variables described in Sec.~\ref{sec:selection}.
The shape and normalization of the remaining MJB are determined from the data.

\begin{figure}
\includegraphics[width=80mm]{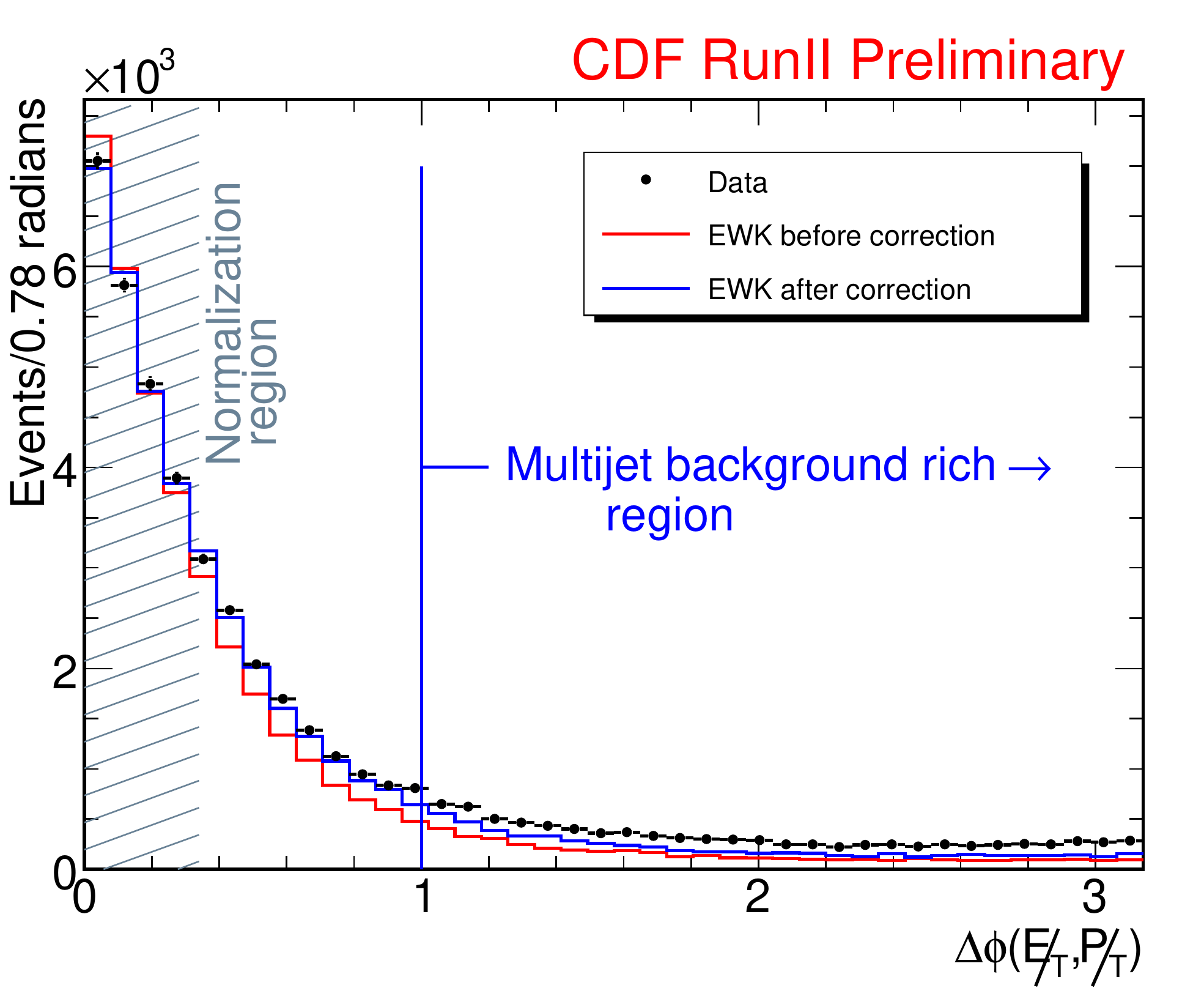}
\caption{Data compared with the sum of the predicted electroweak (EWK) contributions, 
	including the expected signal, for the $\TrackDeltaPhi$ variable defined
	in Sec.~\ref{sec:mjb}.
	The EWK contribution is determined from Monte Carlo, and is corrected
	by normalizing to the data in the peak region, $\TrackDeltaPhi < 0.3$.
	Events with $\TrackDeltaPhi > 1.0$ are used to model the multijet
	background.\label{fig2}}
\end{figure}

To determine the MJB contribution, a region dominated by MJB is isolated in the 
data.  A missing transverse momentum vector $\PtVec$, analogous to the calorimeter-based
$\MetVec$, is constructed from the vector sum of the transverse momenta
of all particles measured in the tracking system.  For $\MET$ arising from 
neutrinos, the $\PtVec$ and $\MetVec$ will usually be closely aligned and the
azimuthal angle $\TrackDeltaPhi$ between them will be small.
The $\PtVec$ is largely uncorrelated to the $\MetVec$ in events where jet 
energies have not been measured accurately; thus the MJB is expected to
be the dominant contribution at large values of $\TrackDeltaPhi$.  The
distribution of $\TrackDeltaPhi$ for events satisfying the selection criteria is shown
in Figure \ref{fig2}.  The EWK Monte Carlo is normalized to the data in the 
peak region of $\TrackDeltaPhi < 0.3$, where the MJB contribution is negligible.
Additional corrections to the Monte Carlo are made based on comparisons of
$Z \to \mu\mu$ Monte Carlo to $Z \to \mu\mu$ events in data.  Because muons are 
minimum ionizing particles which deposit little energy in the calorimeter, 
from calorimeter activity $Z \to \mu\mu$ events appear to have real $\MET$.
This makes $Z \to \mu\mu$ an excellent ``standard candle'' for processes with
$\MET$.

The region of $\TrackDeltaPhi > 1.0$ is dominated by MJB.  Subtracting the 
corrected EWK Monte Carlo from the data in the $\TrackDeltaPhi > 1.0$ region
gives an estimate of the remaining MJB in the signal sample.  We now
determine the shape of the MJB in any variable of interest using events in
the $\TrackDeltaPhi > 1.0$ region, with the EWK contribution subtracted.  
The shapes of the derived MJB distributions 
are also verified with a high-statistics dijet {\sc pythia} Monte Carlo sample.  The
systematic uncertainties associated with this data-driven method of determining 
the MJB contribution come primarily from potential differences between the MJB 
contribution in the $\TrackDeltaPhi < 1.0$ and $\TrackDeltaPhi > 1.0$ regions.
To estimate the size of these uncertainties, we compare distributions in these
regions in the dijet Monte Carlo sample.  The normalization must also be scaled 
up to account for MJB contamination in the region $\TrackDeltaPhi < 1.0$,
and an uncertainty of 20\% is applied to the total MJB yield.

\section{Analysis Technique\label{sec:technique}}
The analysis proceeds via a fit of the signal and background contributions 
to the dijet mass $M_{jj}$ distribution.  The signal extraction is performed 
with an unbinned maximum likelihood fit using the {\sc RooFit} program \cite{roofit}.  
Three $M_{jj}$ template distributions are used in the fit. 
The first is the signal shape.  This template is obtained from a Gaussian plus 
polynomial fit to the signal Monte Carlo $M_{jj}$ distribution, which is a 
combination of the $WW$, $WZ$, and $ZZ$ $M_{jj}$ distributions weighted by the 
predicted SM cross sections.  The mean and width of the Gaussian are linearly 
dependent on the jet energy scale (JES), which is constrained to be in the 
range allowed by external measurements \cite{JES}.
The second template is the EWK shape, a combination of the $M_{jj}$ distributions 
for all backgrounds taken from Monte Carlo as described in Sec.~\ref{sec:ewk}.
The third is the MJB template, which is determined by forming the $M_{jj}$
distribution in events with $\TrackDeltaPhi > 1.0$ as described in Sec.~\ref{sec:mjb}.  
The MJB $M_{jj}$ distribution is modeled by an exponential
which is used as the template in the fit.  Based on studies with the dijet
{\sc pythia} Monte Carlo sample, a 20\% uncertainty is taken on the slope of the 
exponential to account for uncertainties on the shape of the MJB contribution.

In the fit to data, the JES, the MJB normalization, and the slope of
the MJB exponential
enter as Gaussian constraints, allowed to float within the predetermined
uncertainties.  The yield of signal and EWK background events are floating 
in the fit with no constraints.

\section{Systematics\label{sec:syst}}
We address separately two classes of systematic sources, those that
affect the signal extraction and those that affect the signal
acceptance in the cross section calculation.  The signal extraction
systematic uncertainties come from uncertainties on the shapes of
the signal and background templates.  These shape uncertainties
include effects of the jet energy resolution (JER), the jet energy scale 
(JES), and the uncertainties on the shapes of the MJB and EWK background.
The JES and the shape and normalization of the MJB are
treated as nuisance parameters in the fit, with a Gaussian constraint
to their expected values.  Thus these uncertainties are already
accounted for in the statistical uncertainty on the extraction.  The
signal resolution uncertainty is determined by smearing each jet in the
Monte Carlo by the expected uncertainty on the JER.

%
\begin{table}
\caption{The systematic uncertainties described in Sec.~\ref{sec:syst}
	and their effect on the number of extracted signal
	events, the acceptance, and the cross section measurement.  
	All systematics are added in quadrature.  Shape uncertainties
	on the MJB background and JES are included in the statistical
	uncertainty returned by the fit.\label{tab1}}
\begin{tabular}{llc}
\hline\hline
 & Systematic Source & \% Uncertainty \\
\hline
\multirow{2}{*}{\textbf{Extraction}} & EWK background shape & 7.7 \\
 & Resolution & 5.6 \\
\hline 
 & \textbf{Total extraction} & 9.5 \\
\hline\hline
\multirow{6}{*}{\textbf{Acceptance}} & JES & 8.0 \\
 & JER & 0.7 \\
 & $\MET$ resolution model & 1.0 \\
 & Trigger inefficiency & 2.2 \\
 & ISR/FSR & 2.5 \\
 & PDF & 2.0 \\
\hline
 & \textbf{Total acceptance} & 9.0 \\
\hline\hline
 & \textbf{Luminosity} & 5.9 \\
\hline
 & \textbf{Total cross section} & 14.4 \\
\hline\hline
\end{tabular}
\end{table}

The EWK background $M_{jj}$ template is taken from Monte Carlo modeling 
of all non-MJB backgrounds.  The primary EWK backgrounds include a gauge 
boson accompanied by jets.  To determine a shape uncertainty on the EWK
background, an alternative background model is developed using $\gamma$+jets 
data.  There are similarities between $V$+jets and $\gamma$+jets production, 
but the kinematics are not identical due primarily to the large mass 
difference between the $W/Z$ bosons and the photon.
Also, selection requirements applied to the $\gamma$+jets
events are not identical to those used in the $\MET$+jets sample.
The $W/Z$ decays involving neutrinos leave a signature of $\MET$ in the
detector, while the photon's $E_T$ is measured in the calorimeter.  
For this reason, in $\gamma$+jets events we require the 
vector sum of the photon $E_T$ and any $\MET$ in the event to be
greater than 60 GeV, treating this sum as analogous to the $\MET$ in
$V$+jets events.  
To correct for the differences in kinematics and selection, the 
$\gamma$+jets events are weighted by the ratio of the $M_{jj}$ distributions 
in the EWK background Monte Carlo samples to a $\gamma$+jets {\sc pythia}
sample. This method accounts for any production differences, while 
allowing common sources of uncertainty, such as detector effects, parton 
distribution functions, and initial and final state radiation, to cancel out.
Another consideration in the $\gamma$+jets data sample
is the contribution of $\gamma+V$ events, which will cause a peak in
the $M_{jj}$ distribution.  We use a $\gamma+V$ {\sc pythia} Monte Carlo
sample to subtract this contribution.  
After these corrections, the difference between the $\gamma$+jets 
and default EWK background shapes is very small.
We then use the adjusted $\gamma$+jets $M_{jj}$ distribution to
perform a signal extraction fit.  The systematic uncertainty due to the
shape of the EWK background is estimated as the difference in the final
parameter values between the results obtained from this fit and the fit 
with the default EWK background shape.  
This method accounts for a combined effect of JES, JER, and modeling of 
jets in the Monte Carlo.

The dominant source of systematic uncertainty on the signal acceptance
and the cross section measurement is the uncertainty 
associated with the jet energy scale.  The JES affects several of the
variables used in the event selection.  The effect of the JES uncertainty
is quantified by varying the jet energies in the signal Monte Carlo by
the $\pm 1\sigma$ variations of the JES.  Other sources of systematic
uncertainty affecting the measured cross section are the JER, the 
$\MET$ resolution model used in calculating the $\MET$ significance, 
the trigger inefficiency calculated from $Z \to \mu\mu$ events, the 
initial and final state radiation (ISR/FSR), and the parton
distribution functions (PDF).    An additional uncertainty originating from 
the luminosity measurement is also taken into account~\cite{Acosta:2002hx}.
A summary of the sources of systematic uncertainty is given in Table \ref{tab1}.

\section{Results\label{sec:results}}
The measured yields for signal and background from the $M_{jj}$ fit to
data are given in Table \ref{tab2}, with the extracted number of signal
events measured to be $1516\pm239{\rm (stat)}\pm144{\rm (syst)}$.
The acceptances for the $WW$, $WZ$, and $ZZ$ processes are estimated as
2.5\%, 2.6\%, and 2.9\% respectively from the Monte Carlo simulations.
In the calculation of the combined cross section, we assume each signal
contributes proportionally to its predicted SM cross section, which is
11.7 pb for $WW$, 3.6 pb for $WZ$, and 1.5 pb for $ZZ$.
Since the sample utilizes a large number of trigger paths, the 
luminosity of the sample is best calculated by counting the number
of $Z \to \mu\mu$ events in the $\MET$-triggered sample and comparing
to the well understood muon-triggered data.  This gives a total
effective luminosity of 3.5 fb$^{-1}$.  With this information, the
measured signal events correspond to a cross section of
$\sigma(\ppbar \to VV+X) = 18.0\pm2.8{\rm (stat)}\pm2.4{\rm (syst)}\pm1.1{\rm (lumi)}$ 
pb, in agreement with a SM prediction of $16.8\pm0.5$ pb obtained from the
{\sc mcfm} program \cite{mcfm} using the CTEQ6.1M PDFs \cite{cteq}.

\begin{figure}
\includegraphics[width=80mm]{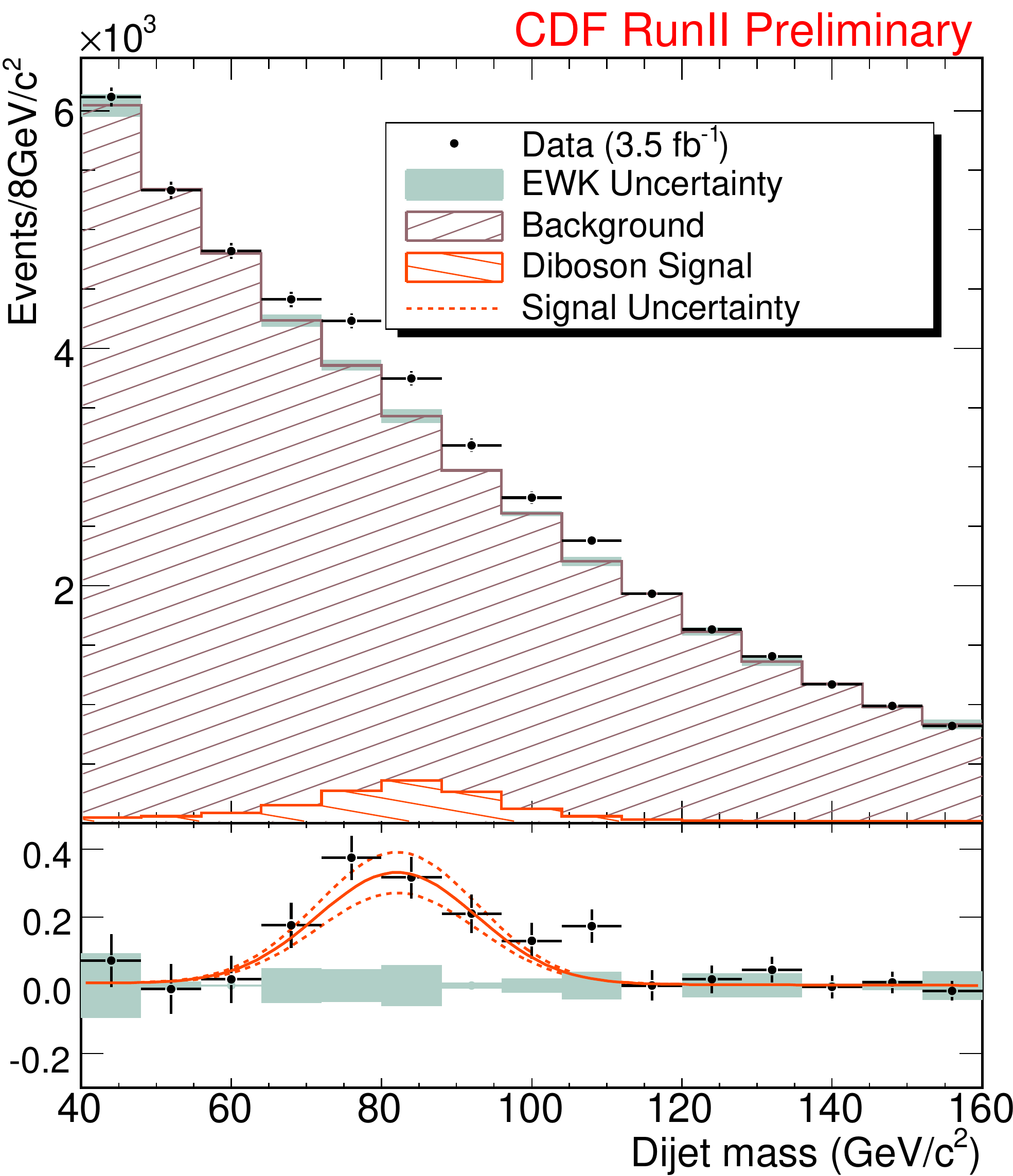}
\caption{Top: Comparison between data and the results of a maximum likelihood fit, 
	with the signal shown unstacked.  The gray band represents the systematic
	uncertainty due to the shape of the EWK background as described in
	Sec.~\ref{sec:syst}.  Bottom: Comparison of the fitted diboson signal
	(solid line) with the background-subtracted data (points).  The
	dashed lines represent the $\pm 1\sigma$ statistical variations
	on the extracted signal, while the gray band again represents
	the systematic uncertainty due to the EWK background shape.\label{fig3}}
\end{figure}

Figure \ref{fig3} shows the fit result and a comparison between the
extracted signal and the data after background subtraction.  With the data
binned as in Figure \ref{fig3} we obtain a $\chi^2$ of 9.4 for 9 degrees
of freedom, corresponding to a p-value of 40\%.
To assess the significance of the observed signal, we consider parameter
variations for the sources of systematic uncertainty.  Then the maximum
likelihood value of a background-only fit is compared to the likelihood
value returned by the full fit, and the difference is converted into
significance numbers.  This method finds that the observed signal 
corresponds to a significance of at least 5.3 standard deviations from 
the background-only hypothesis.

In summary, we use a $\MET$ plus dijet final state to measure the combined
$WW+WZ+ZZ$ cross section in $\ppbar$ collisions at $\sqrt{s} = 1.96$ TeV
to be $18.0\pm2.8{\rm (stat)}\pm2.4{\rm (syst)}\pm1.1{\rm (lumi)}$ pb.
This is consistent with the SM prediction, and is the first observation
in hadronic collisions of the electroweak production of vector boson pairs 
where one boson decays to a hadronic final state.

\begin{table}
\caption{Value of parameters in the model used to fit the $M_{jj}$ distribution.
	Uncertainties shown are statistical only.  Of the 44,910 data events 
	which pass the selection criteria, $1516\pm239$ signal events are
	extracted.  The jet energy scale is also extracted from the fit and
	agrees with the default value of 1.0 as measured from 
	calibrations \cite{JES}.\label{tab2}}
\begin{tabular}{lc}
\hline\hline
Parameter & Fitted Value \\
\hline
Jet energy scale & $0.985\pm0.019$ \\
Yield of EWK background events & $36140\pm1230$ \\
Yield of MJB events & $7249\pm1130$ \\
Yield of diboson candidates & $1516\pm239$ \\
\hline\hline
\end{tabular}
\end{table}

\begin{acknowledgments}
We thank the Fermilab staff and the technical staffs of the participating institutions 
for their vital contributions. This work was supported by the U.S. Department of Energy 
and National Science Foundation; the Italian Istituto Nazionale di Fisica Nucleare; the 
Ministry of Education, Culture, Sports, Science and Technology of Japan; the Natural 
Sciences and Engineering Research Council of Canada; the National Science Council of 
the Republic of China; the Swiss National Science Foundation; the A.P. Sloan Foundation; 
the Bundesministerium f\"ur Bildung und Forschung, Germany; the World Class University 
Program, the National Research Foundation of Korea; the Science and Technology Facilities 
Council and the Royal Society, UK; the Institut National de Physique Nucleaire et Physique 
des Particules/CNRS; the Russian Foundation for Basic Research; the Ministerio de Ciencia 
e Innovaci\'{o}n, and Programa Consolider-Ingenio 2010, Spain; the Slovak R\&D Agency; 
and the Academy of Finland. 
\end{acknowledgments}

\bigskip 

\end{document}